\documentclass[12pt,preprint]{aastex}
\begin{document}
\title{No detectable radio emission from the magnetar-like pulsar in Kes 75}
\author{Anne M. Archibald\altaffilmark{1}}
\email{aarchiba@physics.mcgill.ca}
\author{Victoria M. Kaspi\altaffilmark{1}}
\author{Margaret A. Livingstone\altaffilmark{1}}
\author{Maura A. McLaughlin\altaffilmark{2}}
\altaffiltext{1}{Department of Physics, McGill University, 3600 University St, Montreal, QC H3A 2T8, Canada}
\altaffiltext{2}{Department of Physics, West Virginia University, Morgantown, WV 26505, USA}
\begin{abstract}
The rotation-powered pulsar PSR~J1846$-$0258 in the supernova remnant Kes~75 was recently shown to have exhibited magnetar-like X-ray bursts in mid-2006. Radio emission has not yet been observed from this source, but other magnetar-like sources have exhibited transient radio emission following X-ray bursts. We report on a deep $1.9~\textrm{GHz}$ radio observation of PSR~J1846$-$0258 with the 100-m Green Bank Telescope in late 2007 designed to search for radio pulsations or bursts from this target. We have also analyzed three shorter serendipitous $1.4~\textrm{GHz}$ radio observations of the source taken with the 64-m Parkes telescope during the 2006 bursting period. We detected no radio emission from PSR~J1846$-$0258 in either the Green Bank or Parkes datasets. We place an upper limit of $4.9~\mu\textrm{Jy}$ on coherent pulsed emission from PSR~J1846$-$0258 based on the 2007 November 2 observation, and an upper limit of $27~\mu\textrm{Jy}$ around the time of the X-ray bursts. Serendipitously, we observed radio pulses from the nearby RRAT~J1846$-$02, and place a $3\sigma$ confidence level upper limit on its period derivative of $1.7\times 10^{-13}$, implying its surface dipole magnetic field is less than $2.6\times 10^{13}~\textrm{G}$.
\end{abstract}
\section{Introduction}

Soft Gamma Repeaters (SGRs) and Anomalous X-ray Pulsars (AXPs) are
now well accepted as being different though similar manifestations of
``magnetars'' -- isolated, young ultra-highly magnetized neutron stars
whose radiation is powered by their magnetic field \citep[see][for a review]{wt06}.  However, there remain some outstanding puzzles in the
magnetar picture.  One is the connection between
magnetars and high-magnetic-field rotation-powered pulsars.  There are now known seven
otherwise ordinary rotation-powered pulsars having inferred surface dipole magnetic field
$B>4\times 10^{13}$~G (computed using the standard formula $B=3.2\times 10^{19}~\textrm{G}\sqrt{P\dot P}$; the quantum critical field $B_{\mathrm{QED}} = 4.4 \times
10^{13}$~G).  Most of these high-magnetic-field pulsars, for example PSRs J1718$-$3718, J1847$-$0130 and J1814$-$1744, have $B$ 
similar to or greater than those
measured for \emph{bona fide} magnetars, yet show only faint X-ray emission, if any
\citep{msk+03,km05,pkc00}.  The Rotating Radio Transient (RRAT) J1819$-$1458 also has a field comparable to magnetars, but it exhibits modest X-ray emission \citep{mrg+07}.  \citet{gkc+05} suggest
that another high-B radio pulsar, PSR~J1119$-$6127, shows evidence for
possibly anomalous X-ray emission in the form of a high surface temperature and high pulsed fraction for
thermal emission. 

``Transient'' magnetars, such as XTE~J1810$-$197, are typically X-ray faint but
occasionally have major AXP-like outbursts \citep[e.g.][]{ims+04}.
Similarly, the candidate transient AXP AX~J1845$-$0258 is
either extremely faint or
undetectable in quiescence, but was seen to be at least several hundred
times brighter than usual in a 1993 outburst \citep{vgtg00,tkgg07}.   
\citet{km05} suggest such objects could
be related to high-B radio pulsars, noting the similarity of the spectrum of
the high-B radio pulsar PSR~J1718$-$3718 to that of XTE J1810$-$197 in
quiescence.  This suggestion was supported by the discovery of radio 
pulsations from XTE~J1810$-$197 after its major outburst 
(Camilo et al. 2006), albeit
with an unusual radio spectrum and unusual radio variability properties.
Camilo et al. (2007) report a second magnetar, 1E 1547.0$-$5408, in outburst with similar radio properties.

Very recently, the proposed connection between high-B radio pulsars
and magnetars was given a major boost by the discovery of SGR-like
X-ray bursts and a several-month-long flux enhancement from 
PSR~J1846$-$0258, which was previously
thought to be a purely rotation-powered pulsar \citep{gg+08}.  
This source has a quiescent X-ray luminosity that could be rotation-powered,
and has other properties of rotation-powered pulsars such as a pulsar
wind nebula \citep[PWN; ][]{hcg03,ks08,nsgh08} and an
unremarkable braking index of 2.65$\pm$0.01 \citep{lkgk06}. PSR~J1846$-$0258 has a period of $326~\textrm{ms}$, an estimated dipole magnetic field of $4.9\times 10^{13}~\textrm{G}$, and an estimated spin-down age of 884 years \citep{lkgk06}. Its accurate position, obtained with the \emph{Chandra} observatory, clearly associates it with the supernova remnant (SNR) Kes~75 \citep{hcg03}. 
However, no radio emission has yet been detected from the pulsar \citep{kmj+96}.

Here we report on radio observations of PSR~J1846$-$0258, obtained fortuitously
on the same day as the onset of its observed magnetar-like X-ray behaviour, as well as over a year after this episode.  Using these data, we have searched for coherent radio pulsations.  We also report on our search for
single radio bursts from this source.  We find neither radio pulsations nor
bursts and set upper limits on both.  However we do detect and describe radio bursts from a 
nearby, unrelated Rotating Radio Transient (RRAT) J1846$-$02 \citep{mll+06}. 

\section{Observations and Results}
We obtained a deep observation of PSR~J1846$-$0258 with the 100-m Robert C. Byrd telescope at Green Bank, WV, operated by the NRAO\footnote{The National Radio Astronomy Observatory is a facility of the National Science Foundation operated under cooperative agreement by Associated Universities, Inc.}. The RRAT~J1846$-$02 \citep{mll+06} is estimated to be within two arcminutes of PSR~J1846$-$0258, though its position is uncertain by some seven arcminutes. Since the well-known periods of RRAT~J1846$-$02 ($4.4767~\textrm{s}$) and PSR~J1846$-$0258 ($326.29~\textrm{ms}$) are incommensurable (their ratio is $13.720$), they are clearly distinct sources. We were also able to analyze several archival observations, taken with the 64-m Parkes radio telescope in New South Wales, Australia, covering the key period during and just after the the X-ray bursts described by \citet{gg+08}. Together these observations constrain radio emission associated with the X-ray bursts on both long and short timescales.

There are two distance estimates in the literature for PSR~J1846$-$0258, $21~\textrm{kpc}$ \citep{bh84} and $5-7.5~\textrm{kpc}$ \citep{lt08}. Using the NE2001 free electron density model \citep{cl02}, these distances predict a range of dispersion measures (DMs) from $210~\textrm{pc}~\textrm{cm}^{-3}$ to $1441~\textrm{pc}~\textrm{cm}^{-3}$; the DM through the entire Galaxy in this direction is estimated to be $1464~\textrm{pc}~\textrm{cm}^{-3}$, although these figures are quite uncertain. The interstellar scattering times predicted by the NE2001 model range from $0.03~\textrm{ms}$ to  $17~\textrm{ms}$ for a $1.9~\textrm{GHz}$ observing frequency, and from $0.1~\textrm{ms}$ to $65~\textrm{ms}$ for a $1.4~\textrm{GHz}$ observing frequency. 

The intrinsic radio pulse width of PSR~J1846$-$0258 is uncertain; the X-ray pulse is quite broad \citep{gvbt00}, but X-ray pulse morphology can be very different from radio pulse morphology \citep[e.g. as described in][XTE~J1810$-$197 had a broad X-ray profile but a small radio duty cycle]{gh05,crp+07}. For the (mostly radio) pulsars listed in \citet{mhth05} the average duty cycle was about 5\%, and the transient AXP XTE~J1810$-$197 was observed with a variable duty cycle with typical value of 2\% \citep{crp+07}. Single pulses from the RRATs had lengths from $2$ to $30~\textrm{ms}$ \citep{mll+06}. We elected to focus our search on a duty cycle of about $~1$\%, that is, pulses of length $3~\textrm{ms}$. We ran folding and single-pulse searches (see below) that were sensitive to somewhat shorter pulses, but pulses shorter than the time resolution of our searches would be detected (or not) according to their flux averaged over our time resolution; if the pulse length is $\alpha<1$ times our search's time resolution, our sensitivity to the peak flux is reduced by a factor of $\sqrt{\alpha}$. Our tools are generally fairly sensitive to pulses longer than the time resolution, so we have tried to choose parameters that allow us to detect a broad range of duty cycles around 1\%.

\subsection{Deep single observation with the Green Bank Telescope}
\label{sec:gbtobs}
We observed PSR~J1846$-$0258 while pointing at RA $18^{\mathrm{h}} 46^{\mathrm{m}} 24.96^{\mathrm{s}}$, DEC $-2\degr 58\arcmin 30.72\arcsec$ \citep[J2000,][]{hcg03} for 201 minutes, beginning at MJD 54406.969 (2007 November 2 23:14 UTC), using the Green Bank Telescope. We used the S-band receiver,
observing a bandwidth of 600~MHz centered at 1950~MHz, feeding into the
SPIGOT pulsar backend \citep{k+05}, which recorded 1024 channels (only 768 of which were within our
bandpass) of 16-bit samples with a time resolution of 81.92~$\mu$s. The two linear polarization channels were summed. We performed RFI excision using the program \texttt{rfifind} from the software package PRESTO \citep{ran01, rem02}. RFI conditions were mild, requiring only $\sim 4$\% of the data to be discarded. 

We analyzed the data in three ways: by trial folding, by using a Fourier-domain blind periodicity search, and by searching for bright dispersed single pulses.

Trial folding was carried out using the program \texttt{prepfold}
from PRESTO. We folded the
data into 64 pulse phase bins at 1281 evenly-spaced DMs ranging from zero to $4288~\textrm{pc}~\textrm{cm}^{-3}$.  These DM spacings correspond to a shift of a single profile
bin over the whole observation time per step.  In case the pulsar has a very large or very small duty cycle we also repeated the search with 16 phase bins and 256 phase bins, with corresponding numbers of DM trials and the same minimum and maximum DM. The ephemeris we used for folding was
taken from contemporaneous \textit{Rossi X-ray Timing Explorer} (\textit{RXTE}) observations, part of the program described in \cite{lkgk06}. The timing model we used specifies the pulsar's rotational frequency as
\begin{equation}
\nu(t) = \nu_0 + \dot \nu_0 (t-t_0) + \ddot \nu_0 (t-t_0)^2/2,
\end{equation}
where $\nu_0 = 3.064756108(4)~\textrm{Hz}$, $\dot \nu_0 = -6.6888(4) \times 10^{-11}~\textrm{Hz}~\textrm{s}^{-1}$, $\ddot \nu_0 = 4.81(12) \times 10^{-20}~\textrm{Hz}~\textrm{s}^{-2}$, and $t_0$ is MJD $54376.452350$. Period and period derivative uncertainties obtained from this ephemeris are insignificant over the course of this observation.

We detected no radio pulsations at or near the trial ephemeris.  The GBT observing guide describes a
system temperature $T_{\mathrm{sys}}$ of approximately $19.5~\textrm{K}$ for this band and a gain $G$
of approximately $1.75~\textrm{K}~\textrm{Jy}^{-1}$. Since the band at which we observed, $2~\textrm{GHz}$, is more or less free of both ionospheric and tropospheric effects and the GBT receivers are quite stable, this number should be reasonably accurate. We modeled the background $T_{\mathrm{BG}}$ as a continuum of $400~\textrm{K}$ at $408~\textrm{MHz}$ \citep{hssw82}, which we assumed to have a spectral index of $-2.6$, plus a contribution $S_{\mathrm{SNR}}$ from the supernova remnant of $10~\textrm{Jy}$ at $1.4~\textrm{GHz}$ with a spectral index of $-0.7$ \citep{gre06}. To compute the upper limit on the pulsar's mean flux, $S_{\mathrm{min}}$, we use the expression \citep[e.g.][]{lk05}:
\begin{equation}
\label{eqn:ul1}
S_{\mathrm{min}} = (S/N)_{\mathrm{min}} \beta \sqrt{\frac{W}{P-W}} \sigma ,
\end{equation}
where $(S/N)_{\mathrm{min}}$ is the signal-to-noise threshold at which we would certainly have detected the pulsations, $\beta$ is a unitless factor describing quantization losses, $P$ is the pulsar's period, $W$ is the time per period during which the pulsar is on, and $\sigma$ is the noise RMS amplitude, given by:
\begin{equation}
\label{eqn:ul2}
\sigma = \frac{(T_{\mathrm{sys}}+T_{\mathrm{BG}})/G+S_{\mathrm{SNR}}}{\sqrt{n_p t \Delta f}},
\end{equation}
where $n_p$ is the number of polarizations added (two in our case), $t$ is the observation length (201 minutes), and $\Delta f$ is the bandwidth ($600~\textrm{MHz}$).
Since the Spigot uses a three-level digitzer, we used $\beta=1.16$ \citep{lk05}. For this calculation, we assumed the duty cycle was $1/64$, and that we would have detected any pulsar with a peak more than four times the RMS noise per bin.  This gives an upper limit of $4.9~\mu\textrm{Jy}$. Longer duty cycles give higher upper limits, up to $43~\mu\textrm{Jy}$ (assuming a square-wave profile).

We carried out the blind periodicity search using the program \texttt{accelsearch}, again from the PRESTO toolkit. We produced a dedispersed time series with time resolution $0.32768~\textrm{ms}$ for $2000$ dispersion measures spaced by $2~\textrm{pc}~\textrm{cm}^{-3}$ from zero to $4000~\textrm{pc}~\textrm{cm}^{-3}$. We searched only for unaccelerated pulsations, summing up to sixteen harmonics (the maximum supported by \texttt{accelsearch}). We detected no pulsations.  We examined all candidates with a signal-to-noise ratio above $6$. Applying Equations \ref{eqn:ul1} and \ref{eqn:ul2} as we did for our coherent pulsation search, and assuming a duty cycle of $1/32$ (since the use of sixteen harmonics smears the folded profile by this much) we place an upper limit on coherent pulsations at any frequency of $10~\mu\textrm{Jy}$. 

\begin{figure}
\centering
\plotone{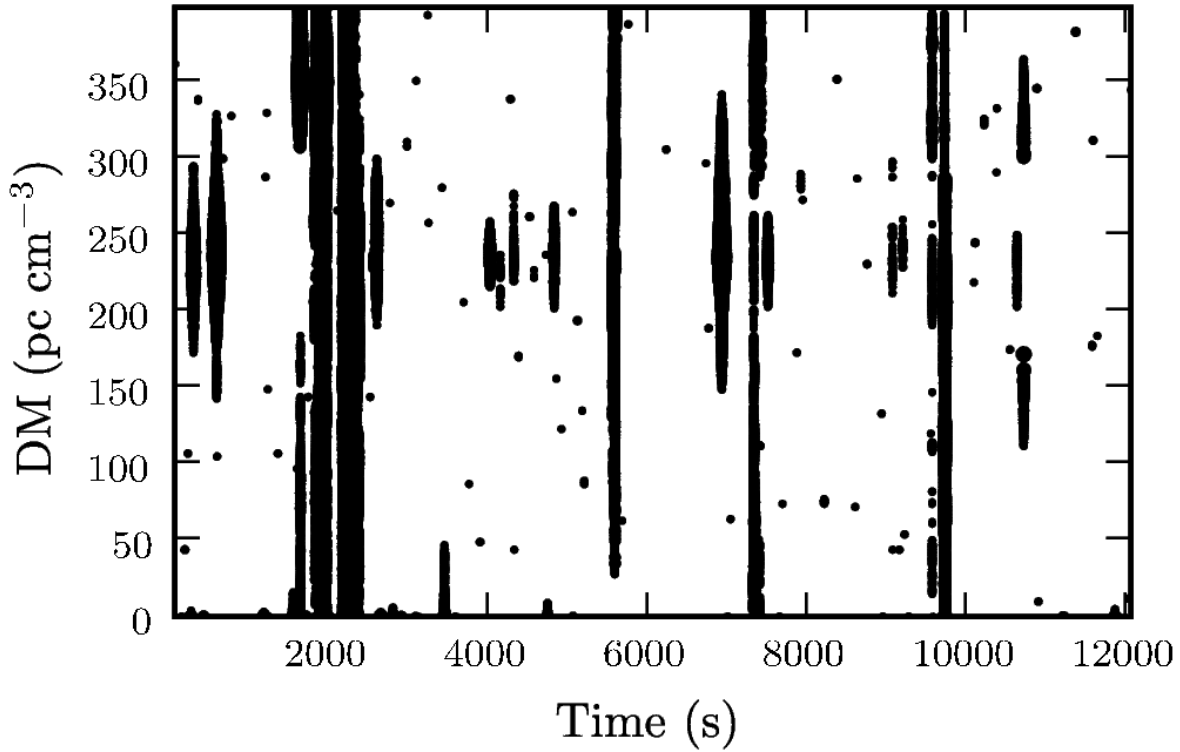}
\caption{
Single-pulse search results plot from our three-hour Green Bank Telescope observation. Each single pulse detection above a threshold of $6\sigma$ is plotted as a circle whose diameter indicates the significance of the detection. Note the series of pulses from the RRAT~J1846$-$02 around a DM of $237~\textrm{pc}~\textrm{cm}^{-3}$. The final, brightest, pulse from the RRAT, just before $11000~\textrm{s}$, appears only below a DM of $170~\textrm{pc}~\textrm{cm}^{-3}$ and above DM $300~\textrm{pc}~\textrm{cm}^{-3}$ because it was so strong it was erroneously identified as a receiver gain change in dedispersed time series closer to the correct DM. Vertical groups of strong detections extending down to zero DM are RFI.
}\label{fig:singlepulse}
\end{figure}

\begin{figure}
\centering
\plotone{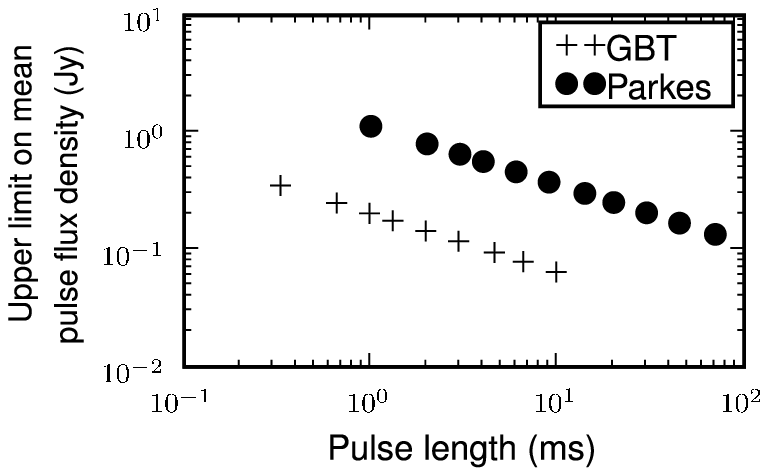}
\caption{
Upper limits on the mean flux density for single pulses of different lengths for our GBT (Sec.~\ref{sec:gbtobs}) and Parkes (Sec.~\ref{sec:pkobs}) observations.
}\label{fig:spupper}
\end{figure}

We carried out the single pulse search using \texttt{single\_pulse\_search.py}, again from the PRESTO toolkit. This software operates on the same dedispersed time series described above and computes running boxcar averages at a range of widths from our dedispersed sample time of $0.32768~\textrm{ms}$ up to $9.8304~\textrm{ms}$. Statistically significant detections are recorded (after some automated sifting to reduce the effects of radio frequency interference and receiver gain changes), and the results are plotted in Figure~\ref{fig:singlepulse}. The peak-flux threshold for detection of single pulses depends on the length of the pulses; see Figure~\ref{fig:spupper}. The single-pulse searching code we used removes certain strong pulses which it interprets as receiver gain changes. As a result, the last and brightest pulse was erroneously removed from the list of single-pulse detections between a DM of $170~\textrm{pc}~\textrm{cm}^{-3}$ and a DM of $300~\textrm{pc}~\textrm{cm}^{-3}$. It was nevertheless present in our dedispersed time series and we used it in our timing analysis.


The signature of a single bright astrophysical pulse should be a collection of single-pulse detections well above the $\textrm{DM}=0$ axis. Indeed we find a number of such pulses, at a DM of approximately $237~\textrm{pc}~\textrm{cm}^{-3}$. Closer examination reveals twelve bright single pulses, all clustered around this DM. A number of other pulse detections were observed, but all appear to be either terrestrial RFI (groups of detections at the same time and strongest at zero DM) or noise.

When we fold the arrival times of all single-pulse candidates (apart from certain obvious RFI) according to our ephemeris for the X-ray pulsar in Kes 75, we find that they fall at random phases; in fact a Kuiper test gives a probability of 0.22 that a uniform distribution would give rise to arrival times more unevenly distributed in phase than this. However, the RRAT~J1846$-$02, described in \citet{mll+06}, falls within the GBT's $7'$ beam when it is pointed at PSR~J1846$-$0258. The DM reported in \citet{mll+06} is $239~\textrm{pc}~\textrm{cm}^{-3}$, which closely matches that of our single pulses. Moreover, if we fold the single-pulse arrival times at the reported period of the RRAT, $4.476739(6)~\textrm{s}$, we find that they fall within four milliperiods of the same phase. We therefore infer that the only significant cosmic single pulses in our observation are from RRAT~J1846$-$02. 

Since we detected twelve bright pulses from the RRAT~J1846$-$02 over the course of approximately three hours, we can compute a period for the RRAT based on the timing of single pulses. We selected the single brightest pulse, smoothed it by convolution with a von Mises distribution \citep[e.g.][]{mard75} of full width at half maximum $10~\textrm{ms}$, and used it as a template. The barycentric arrival time of each pulse was estimated by Fourier-domain cross-correlation with this template. We then used the published period to compute the number of turns between each pair of pulses, and adjusted the period and starting phase to minimize the root-mean-squared residual phase. 

While we would expect rather small formal errors on the arrival times for data of this quality, we do observe about $4~\textrm{ms}$ residual jitter, possibly due to pulse shape variations. We have used the RMS variation of the residuals as an estimate of the uncertainty on each arrival-time measurement. Taking this into account, we obtain a barycentric period estimate of $4.4767435(2)~\textrm{s}$ at epoch MJD 54407. Subtracting this from the reported period of $4.476739(3)~\textrm{s}$ \citep[epoch MJD 53492,][]{mll+06} and dividing by the elapsed time gives a period derivative estimate of $(5.5\pm 3.8)\times 10^{-14}$, implying a $3\sigma$ upper limit on the period derivative of $1.7\times 10^{-13}$. Using $B=3.2\times 10^{19}~\textrm{G}\sqrt{P\dot P}$ this implies an upper limit on the surface dipole magnetic field of $2.6\times 10^{13}~\textrm{G}$.

\subsection{Observations with the Parkes telescope near the bursting epoch}
\label{sec:pkobs}
As part of a program to monitor RRAT~J1846$-$02, three one-hour observations including both it and PSR~J1846$-$0258 were taken at MJDs 53886.64, 53923.54, and 53960.45. Bursts from PSR~J1846$-$0258 were detected in X-ray observations at MJD 53886.92--53886.94 and 53943.46 \citep{gg+08}; fortuitously, one radio observation was taken only six hours before the first of these bursts.

The radio observations were taken with a 256 MHz bandwidth centered at 1390 MHz. Each observation was 60 minutes long, and was recorded with the centre beam of the Parkes Multibeam receiver. They were one-bit digitized with a time resolution of 0.1 ms and 512 spectral channels. We analyzed them in the same three ways as for the GBT data --- folding at the known period, blind periodicity searching, and single-pulse searching --- using the same software tools. Unfortunately, because of the timing anomaly that the source underwent around this time \citep{gg+08}, we were not able to produce a phase-coherent timing solution from contemporaneous \textit{RXTE} data. We were able to obtain period estimates from periodograms of the \textit{RXTE} data, and we searched a range of periods in the radio data to be certain of including the true period. 

We folded each observation using 64 phase bins at 129 periods centered at the peridogram frequency of the nearest \textit{RXTE} observation and spaced over $1.4~\mu\textrm{s}$, and 1025 DMs spaced from $0$ to $4008~\textrm{pc}~\textrm{cm}^{-3}$.  As before we also folded with 16 and 256 bins.  The center frequencies we used for this folding are $3.067684~\textrm{Hz}$, $3.067457~\textrm{Hz}$, and $3.067240~\textrm{Hz}$, from RXTE observations at MJDs $53886.91$, $53920.95$, and $53955.57$ for the Parkes observations on MJD $53886.64$, $53923.54$, and $53960.45$ respectively.  We saw no coherent pulsations.  Using the same background figures and duty cycle assumptions as above, and telescope parameters obtained from the Parkes Radio Telescope Users' Guide ($T_{\mathrm{sys}}=23.5~\textrm{K}$, $G=0.67~\textrm{K}~\textrm{Jy}^{-1}$, $\beta=1.25$), we estimate a flux upper limit of $27~\mu\textrm{Jy}$ for each observation.

Our blind searches were carried out as above, and we set an upper limit of $\sim 58~\mu\textrm{Jy}$ on all coherent pulsations from the source.  For single-pulse search upper limits, see Figure~\ref{fig:spupper}. We detected single pulses from RRAT~J1846$-$02 in one of these data sets.  Only one pulse in our three-hour GBT observation was brighter than our threshold for detection in the Parkes observations, so a single detection in the three one-hour Parkes observations is not unexpected.

\section{Discussion}
We did not detect radio emission from PSR~J1846$-$0258, in spite of a deep GBT observation and a contemporaneous timing solution.  Our upper limit of $4.9~\mu\textrm{Jy}$ is a substantial improvement over that published in \citet{kmj+96}, which quotes an upper limit at $1520~\textrm{MHz}$ of $100~\mu\textrm{Jy}$.

Two distance estimates for the pulsar/supernova remnant system are found in the literature: $21~\textrm{kpc}$ \citep{bh84}, and more recently $5.1~\textrm{kpc}$--$7.5~\textrm{kpc}$ \citep{lt08}. This newer result is based on more recent HI and $^{13}$CO maps. The smaller distance yields a smaller diameter for the remnant, consistent with the indications that PSR~J1846$-$0258 is very young \citep{lkgk06}. The smaller distance also yields a smaller X-ray luminosity for the pulsar, which had previously appeared to be unusually high \citep[$4.1\times 10^{35}~\textrm{erg}~\textrm{s}^{-1}$, second only to the Crab]{hcg03}. We will assume this more recent measurement is correct.

The transient AXP XTE~J1810$-$197 was detected with $1.4~\textrm{GHz}$ radio flux densities peaking at $1500~\textrm{mJy}$ and fading to $10~\textrm{mJy}$ over the course of about 6 months \citep{c+07}, with a flat enough spectrum that flux densities at $1.4~\textrm{GHz}$ and $1.9~\textrm{GHz}$ are comparable \citep{crh+06}. Its estimated distance is $3.5~\textrm{kpc}$ \citep{c+07}. If the source we observed had the same luminosity at the far end of the \citet{lt08} estimated distance range, $7.5~\textrm{kpc}$, we would expect a flux density ranging from $320~\textrm{mJy}$ down to $2~\textrm{mJy}$.  The transient AXP 1E~1547.0$-$5408 was observed to have a flux density $\sim 3~\textrm{mJy}$ at $1.4~\textrm{GHz}$ \citep{crhr07}, rising with frequency; the source is at an estimated distance of $9~\textrm{kpc}$ \citep{crhr07}. If PSR~J1846$-$0258 had the same luminosity we would expect a flux of $4~\textrm{mJy}$.  All of these figures are significantly higher than our GBT detection threshold of $4.9~\mu\textrm{Jy}$, so we conclude that if PSR~J1846$-$0258 was emitting radio pulsations during our observations, they must either be much weaker than those observed from transient magnetars, or they must be beamed elsewhere. A recent paper by \citet{nsgh08} computes an angle of $60\degr$ between the line of sight and the spin axis of the pulsar, and they combine this with a tentative suggestion  that in order to obtain the observed braking index the magnetic inclination should be approximately $9\degr$ \citep{mela97,lkgk06}. This would imply that our line of sight is, at its closest, $54\degr$ from the magnetic pole, making it unlikely that we would be in the pulsar beam. However, in light of the evidence given by \citet{c+07b} that the X-ray and radio emission beams of XTE~J1810$-$197 are nearly parallel, the fact that we see X-ray emission from PSR~J1846$-$0258 indicates that if it behaved like XTE~J1810$-$197, any radio emission would likely be beamed in our direction as well.  

Thus it appears likely that if PSR~J1846$-$0258 is emitting radio pulsations, they are much weaker than those emitted by the known radio-emitting transient AXPs.  However, since our GBT observation was taken some eighteen months after the X-ray bursting activity, it is possible that the radio emission had faded by the time of our observation. The upper limits obtained from our Parkes observations --- $21~\mu\textrm{Jy}$ --- are also much less than we would have expected to observe from the known transient AXPs, so we can also constrain the brightness for the first two months after the X-ray bursts began.  In particular, the X-ray bursts (four were observed in a one-hour observation, taking place only six hours after the first of our Parkes observations) are probably not accompanied by radio bursts. 

Leaving aside the AXP-like behavour of PSR~J1846$-$0258, its radio emission appears to be very faint. The young pulsar PSR~J0205+6449 in the supernova remnant 3C58 is detected in X-rays and very faintly at radio wavelengths. It is at an estimated distance of $3.2~\textrm{kpc}$. The radio flux at $1.4~\textrm{GHz}$ is $45~\mu\textrm{Jy}$ with a spectral index of $-2.1$ \citep{csl+02}. If PSR~J1846$-$0258 had the same luminosity and spectral index we would have received a flux of $4.3~\mu\textrm{Jy}$ at $1.9~\textrm{GHz}$, just comparable to our upper limit. On the other hand, the X-ray luminosity of PSR~J0205+6449 is estimated to be $2.84\times 10^{33} \textrm{erg}~\textrm{s}^{-1}$ \citep{ms+02}, while the X-ray luminosity of PSR~J1846$-$0258 is estimated to be $7\times 10^{34} \textrm{erg}~\textrm{s}^{-1}$ \citep[adjusted to the Leahy et al. distance estimate]{ms+07}. Thus in spite of having an X-ray luminosity more than $20$ times that of PSR~J0205+6449, PSR~J1846$-$0258 appears to have a smaller radio luminosity than PSR~J0205+6449. Beaming may account for this difference.  More generally, the pseudoluminosity limit we set is $0.2~\textrm{mJy}~\textrm{kpc}^2$. Only eighteen pulsars have been detected at radio wavelengths that are fainter than this limit; their luminosities range down to $30~\mu\textrm{Jy}~\textrm{kpc}^2$ \citep{mhth05}. 

It is possible that PSR~J1846$-$0258 produced radio emission that peaked several months after the X-ray event and faded by the time of our GBT observation. XTE~J1810$-$197 was observed to brighten in radio about a year after its X-ray brightening, and it faded over the course of about a year \citep{c+07b}. We intend to search other observations from the monitoring program of RRAT~J1846$-$02 for pulsations coming from PSR~J1846$-$0258.

By comparing the period of RRAT~J1846$-$02 measured from our observation with the published period, we were able to place a $3\sigma$ upper limit on the spin-down rate of $1.7\times 10^{-13}$. 
This gives a $3\sigma$ upper limit on the surface dipole magnetic field of $2.6\times 10^{13}~\textrm{G}$, smaller than any known magnetar, and less than those of 24 more highly magnetized rotation-powered pulsars \citep{mhth05}, including PSR~J1846$-$0258.
We hope to combine this relatively long observation with a timing program being carried out by McLaughlin et al. to yield a phase-coherent timing solution for the RRAT~J1846$-$02.

\section*{Acknowledgements}
\acknowledgements{
The authors thank Scott Ransom for helpful conversations during the writing of this paper. The authors are also grateful to the Parkes Multibeam Pulsar Survey team for providing the Parkes observations.
Support for this work was provided to VMK by an NSERC Discovery Grant
(Rgpin 228738-03), an R. Howard Webster Foundation Fellowship of
the Canadian Institute for Advanced Research, Les Fonds de la Recherche
sur la Nature et les Technologies, a Canada Research Chair and
the Lorne Trottier Chair in Astrophysics and Cosmology. MAL is a Natural Sciences and Engineering Research Council (NSERC) PGS-D Fellow. MAM is supported by a WV EPSCoR grant.
}

\textit{Facilities:} \facility{GBT (S-band receiver)}, \facility{Parkes (multibeam receiver)}

\end{document}